\documentclass [useAMS,usenatbib,usegraphicx]{mn2e}
\usepackage{amssymb}
\usepackage{times}
\title[Large-scale motion of RFGC galaxies in curved space-time]
{Large-scale collective motion of RFGC galaxies in curved space-time}
\author[S. L. Parnovsky and A. S. Parnowski]{S. L. Parnovsky$^1$\thanks{E-mail:
par@observ.univ.kiev.ua} and A. S. Parnowski$^2$ \\
$^1$Astronomical Observatory of Kyiv Taras Shevchenko National University, Observatorna str., 3, 04058, Kyiv, Ukraine \\
$^2$Space Research Institute of NASU \& NSAU, prosp. Akad. Glushkova, 40, korp. 4/1, 03680 MSP, Kyiv-187, Ukraine}
\newlength{\tmpla}
\newlength{\tmplb}
\newlength{\tmplc}
\renewcommand{\vec}{\bmath}
\newcommand{\tens}{\mathbfss}
\begin{document}
\date{Accepted . Received ; in original form \today}
\pagerange{\pageref{firstpage}--\pageref{lastpage}} \pubyear{2010}
\maketitle
\label{firstpage}
\begin{abstract}
We consider large-scale collective motion of flat edge-on spiral galaxies from
the Revised Flat Galaxy Catalogue (RFGC) taking into account the curvature of
space-time in the Local Universe at the scale $100h^{-1}\,\rmn{Mpc}$. We analyse
how the relativistic model of
collective motion should be modified to provide the best possible values of
parameters, the effects that impact these parameters and ways to mitigate them.
Evolution of galactic diameters, selection effects, and difference between
isophotal and angular diameter distances are inadequate to explain this impact.
At the same time, measurement error in \mbox{H\,{\sc i}} line widths and
angular diameters can easily provide such an impact. This is illustrated in a
toy model, which allows analytical consideration, and then in the full model
using Monte Carlo simulations. The resulting velocity field is very close to
that provided by the non-relativistic model of motion. The obtained bulk flow
velocity is consistent with $\Lambda$CDM cosmology.
\end{abstract}
\section{Introduction}

At present time the Universe is essentially inhomogeneous on the scales of
about 10--100 Mpc. The development of initial fluctuations led to an observable
large-scale structure. The regions with increased matter density provide an
additional attraction of surrounding galaxies. The regions with decreased
density, e.g.  voids, also make an input to the collective large-scale motion
of galaxies on the background of Hubble expansion. Investigation of such motion
on one side allows to map the matter density, including dark matter, in the
Local Universe, and on the other side its parameters are linked with
cosmological parameters.  All of this makes the study of collective galaxy
motions important.

In recent years a number of articles was published claiming
that typical velocities of large-scale collective motions are inconsistent with
the standard $\Lambda$CDM model. For example, \citet{ref:WFH09} obtained the
value $407 \pm 81\,\rmn{km\,s}^{-1}$ at the scale $100h^{-1}\,\rmn{Mpc}$,
whereas the $\Lambda$CDM model gives about $250\,\rmn{km\,s}^{-1}$. However, our
estimation of $210 \pm 86\,\rmn{km\,s}^{-1}$ at the same scale, obtained in the
article \citep{ref:APSS09}, is consistent with the $\Lambda$CDM predictions.
Additionally, in the same article we obtained from the peculiar velocities the
constraints on the cosmological parameters $\Omega_m$ and $\sigma_8$ and their
combinations, which match the other more precise constraints like baryonic
acoustic oscillations or WMAP observations.

In the article \citep{ref:APSS09} we used a sample of RFGC galaxies with
measured redshifts and \mbox{H\,{\sc i}} line widths. The Revised Flat Galaxy Catalogue
(RFGC) \citep{ref:RFGC} and its previous version Flat Galaxy Catalogue (FGC)
\citep{ref:FGC} contain the information about Right Ascension and Declination
for the epochs J2000.0 and B1950.0, galactic longitude and latitude, major and
minor blue and red diameters in arcminutes in the POSS-I diameter system,
morphological type of the spiral galaxies according to the Hubble
classification, index of the mean surface brightness and some other parameters,
which are not used in this article. The RFGC contains data about 4236 flat
edge-on spiral galaxies, almost uniformly covering the celestial sphere and
satisfying the conditions $a_b/b_b\ge7$ and $a_b>0\farcm 6$. Here $a_b$ and $b_b$
are the major and minor axial diameters in the $a_{25}$ system. The original
goal of this catalogue was to estimate the distance to galaxies according to
the Tully-Fisher relation in the ``\mbox{H\,{\sc i}} line width -- linear diameter'' version
without using their redshifts. The data about the redshifts and \mbox{H\,{\sc i}} line
widths or gas rotation velocities $V_{rot}$ were taken from different sources.
There were a number of gradually increasing samples of galaxies with such data
\citep{ref:K00,ref:Par01,ref:ParTug04}. The latest version of this sample
including 1623 galaxies was compiled and described by \citet{ref:APSS09}. A
list of peculiar velocities based upon this list in the non-relativistic model
of motion was assembled by \citet{ref:arxiv09}.

In this article we use the same sample, but with different model of collective
motion of galaxies \citep{ref:KudAlex02,ref:KudAlex04}, based upon the general
theory of relativity (GTR). This model was applied earlier to the
previous version of the sample by \citet{ref:ParGayd05}; however, the present
article offers a much more in-depth analysis.

\section{Description of models of collective motion of galaxies}

We assume that there is a three-dimensional velocity field of collective
galactic motions on the background of Hubble expansion. We consider the
galaxies in the sample not as massive objects, but rather as test particles,
whose peculiar velocities are indicators of the large-scale velocity field.
Using the multipole decomposition of large-scale velocity field up to quadratic
terms in distance and switching to the radial component we obtain the
expression for radial velocity of an individual galaxy. The actually measured
radial velocities differ from those predicted by this model due to
deviations from the statistical Tully-Fisher relation, influence of
motions with scales much less than the sample depth (fall towards nearby
attractors) and random errors. Treating these errors as stochastic we can use
the least squares method to calculate the parameters of the model.

\subsection{Non-relativistic model of collective motion}

Before discussing the relativistic model of collective motion, let us briefly
recall the non-relativistic models, introduced in the article \citep{ref:Par01}
and used by \citet{ref:ParTug04} and \citet{ref:APSS09}. We will start from the
more complex of them, namely the DQO-model.  \begin{equation}\label{eqn:DQO}
V=R+V^{dip}+V^{qua}+V^{oct}+\delta V.  \end{equation} Here $V$ is a radial
velocity of the galaxy in the CMB isotropy frame, $R=Hr$ is the Hubble
velocity, $r$ is the distance to the galaxy, $\delta V$ is a random error,
$V^{dip}$, $V^{qua}$ and $V^{oct}$ are the dipole (D), quadrupole (Q) and
octopole (O) components of the non-Hubble cosmic flow. They are given by the
following expressions:
\begin{equation}\label{eqn:D}
\begin{array}{l}
V^{dip}=D_{i}n_{i},
\end{array}
\end{equation}
\begin{equation}\label{eqn:Q}
\begin{array}{l}
V^{qua}=RQ_{ik}n_{i}n_{k}=R\left(q_1(n_1^2-n_3^2)+q_2(n_2^2-n_3^2)\right.\\
\phantom{V^{qua}=}\left.{}+q_3n_1n_2+q_4 n_1 n_3+q_5 n_2
n_3\right),
\end{array}
\end{equation}
\begin{equation}\label{eqn:O}
\begin{array}{l}
V^{oct}=R^{2}(O_{ikl}n_{i}n_{k}n_{l}+P_{i}n_{i})=R^2\left(P_{i}n_{i}\right.\\
\phantom{V^{oct}=}\left.{}+o_1(3n_1n_2^2-n_1^3)+o_2(3n_1n_3^2-n_1^3)\right.\\
\phantom{V^{oct}=}\left.{}+o_3(3n_2n_1^2-n_2^3)+o_4(3n_2n_3^2-n_2^3)\right.\\
\phantom{V^{oct}=}\left.{}+o_5(3n_3n_1^2-n_3^3)+o_6(3n_3n_2^2-n_3^3)\right.\\
\phantom{V^{oct}=}\left.{}+o_7n_1n_2n_3\right).
\end{array}
\end{equation}
Here we use the Einstein rule -- summation by repeated indices; $n_i$ are
Cartesian components of the unit vector $\vec{n}$ towards the galaxy, connected
with the galactic coordinates $l$ and $b$ in the following way:
\begin{equation}\label{eqn:lb}
\begin{array}{l}
n_1=n_z=\sin b,\\
n_2=n_x=\cos l\cos b,\\
n_3=n_y=\sin l\cos b.
\end{array}
\end{equation}
The dipole component (bulk motion) is described by the vector
$\vec{D}$. The quadrupole component is described by the symmetrical traceless
tensor $\tens{Q}$. It has 5 independent parameters $q_i$. The octopole
component can be described by one rank 3 tensor, but we divide it into a trace
characterized by vector $\vec{P}$ and a tensor $\tens{O}$, which is
antisymmetrical with respect to each pair of indices. The latter has 7 independent
parameters $o_i$.

Hubble velocity is determined from the generalized Tully-Fisher relation in the
``angular diameter -- \mbox{H\,{\sc i}} line width'' version by the following formula
\begin{equation}\label{eqn:TF}
\begin{array}{l}
R=(C_1+C_2B+C_3BT+C_4U)\frac{W}{a}\\
\phantom{R=}{}+C_5\left(\frac{W}{a}\right)^2+C_6\frac{1}{a},
\end{array}
\end{equation}
where $W$ is a corrected \mbox{H\,{\sc i}} line width in $\rmn{km\,s}^{-1}$ measured at $50$ per
cent of the maximum, $a$ is a corrected major galaxies' angular diameter in
arcminutes on red POSS and ESO/SERC reproductions, $U$ is a ratio of major
galaxies' angular diameters on red and blue reproductions, $T$ is a
morphological type indicator ($T=I_{t}-5.35$, where $I_{t}$ is a Hubble type;
$I_{t}=5$ corresponds to type Sc), and $B$ is a surface brightness indicator
($B=I_{SB}-2$, where $I_{SB}$ is a surface brightness index from RFGC;
brightness decreases from I to IV). Note that the statistical significance of
each term in eq. (\ref{eqn:TF}) is greater than $99$ per cent according to the
F-test \citep{ref:F,ref:H}.

Thus, the DQO-model contains 24 parameters, namely 3 components of the vector
$\vec{D}$, 6 coefficients $C_i$, 5 parameters $q_i$ of the tensor $\tens{Q}$, 3
components $p_i$ of the vector $\vec{P}$ and 7 parameters $o_i$ of the tensor
$\tens{O}$. By rejecting $V^{oct}$ we get a simpler DQ-model with 14
components. Further rejecting $V^{qua}$ leads to the simplest D-model with 9
components.

\subsection{Relativistic model of collective motion}

The existence of multipole components is due to density inhomogeneities in the
Universe. For homogeneous isotropic Universe instead of equation (\ref{eqn:DQO})
we would have the simple Hubble law $V=Hr$. Hubble expansion is due to cosmology
and thus is most adequately described in the framework of GTR. Such a description
raises a problem of distance measurement. The main types of distances used are
the redshift distance, photometric distance and angular diameter distance. They
are related to each other through formulae containing cosmological parameters.
In our case the natural choice is the angular diameter distance, since we
determine distances by the generalized Tully-Fisher relation using angular
diameters. Thus, when speaking about distance $r$ in relativistic models, we
will mean angular diameter distance.

For the homogeneous isotropic
cosmological models the dependence of the velocity $V=cz$ on $R=Hr$ in the next
order in $R$ has the form
\begin{equation}\label{eqn:R}
V=R+\gamma R^2.
\end{equation}
The coefficient $\gamma$ is expressed through the deceleration parameter $q$ by
\begin{equation}\label{eqn:g}
\gamma=\frac{3+q}{2c}.
\end{equation}
For different types of distance another expression for $\gamma$ should be used.
For the standard $\Lambda$CDM cosmology we have
\begin{equation}\label{eqn:q}
q=\frac{\Omega_m}{2}-\Omega_{\Lambda}=-0.61,
\end{equation}
where $\Omega_m$ and $\Omega_{\Lambda}$ are relative densities of matter,
including dark matter, and dark energy respectively. Numerical estimations are
based on the results of 7-year WMAP observations \citep{ref:WMAP7}.  Introducing
(\ref{eqn:q}) into (\ref{eqn:g}) we obtain
\begin{equation}\label{eqn:g0}
\gamma_0=3.98\cdot 10^{-6}\,\rmn{s\,km}^{-1}.
\end{equation}

Now let us consider a relativistic model of galaxy motion in inhomogeneous
space-time. It was developed by \citet{ref:KudAlex02,ref:KudAlex04}. Instead of
(\ref{eqn:DQO}) we use the equation
\begin{equation}\label{eqn:DQOR}
V_{rel}=R+V^{dip}+V^{qua}+V^{oct}+\gamma R^2+\delta V.
\end{equation}
Here $V_{rel}$ is still given in the CMB frame, the expressions for the dipole
(\ref{eqn:D}) and qudrupole (\ref{eqn:Q}) components remain the same, and the
octopole component assumes the form
\begin{equation}\label{eqn:OR}
V_{rel}^{oct}=R^{2}\left(P_in_i+O_{ijk}n_in_jn_k+S_{ij}n_in_j\right)
\end{equation}
Here $\tens{S}$ is a symmetric traceless tensor, characterized by 5 independent
parameters $s_i$:
\begin{equation}\label{eqn:S}
\begin{array}{l}
S_{ij}n_in_j=s_1(n_1^2-n_3^2)+s_2(n_2^2-n_3^2)\\
\phantom{S_{ij}n_in_j=}{}+s_3n_1n_2+s_4n_1n_3+s_5n_2n_3.
\end{array}
\end{equation}
The parameter $\gamma$ is related to the deceleration parameter by
\begin{equation}\label{eqn:gR}
\gamma=\frac{3+q}{2c}+\frac{1}{3c}Q_{ik}Q^{ik}.
\end{equation}
This equation reduces to equation (\ref{eqn:g}) in absence of quadrupole
component, for example, for homogeneous models. Similarly to the
non-relativistic case, we can reduce the relativistic DQO-model to DQ- and
D-models by rejecting highest-order multipoles.

In the papers \citep{ref:KudAlex02,ref:KudAlex04} it was shown that the relation
\begin{equation}\label{eqn:Weyl}
\begin{array}{l}
C_{\alpha\beta\gamma\delta}V^\beta V^\delta
=\frac{H^2}{c^2}\left(-2cS_{\alpha\gamma}+6Q_{\alpha\gamma}
-3Q^\epsilon_\alpha Q_{\epsilon\gamma}\right.\\
\phantom{C_{\alpha\beta\gamma\delta}V^\beta V^\delta=}\left.
{}-(V_\alpha V_\gamma -g_{\alpha\gamma}Q_{\epsilon\eta}Q^{\epsilon\eta})\right)
\end{array}
\end{equation}
must hold. Here Greek indices denote four-dimensional components,
$g_{\alpha\beta}$ is a metrical tensor, and $C_{\alpha\beta\gamma\delta}$ is a
Weyl tensor. The spatial parts of four-dimensional tensors $Q_{\alpha\beta}$
and $S_{\alpha\beta}$ coincide with three-dimensional tensors $\tens{Q}$ and
$\tens{S}$, and their temporal and mixed parts are much smaller. The
four-dimensional velocity vector $V_\alpha$ can be assumed equal to
$(g_{00}^{-1/2},0,0,0)$.

Now all that remains is to supply the relativistic model with an expression for
the angular diameter distance. Since we consider the terms proportional to
$R^2$ separately, we should remove the terms quadratic in distance from the
generalised Tully-Fisher relation (\ref{eqn:TF}):
\begin{equation}\label{eqn:TFR}
R=(C_1+C_2B+C_3BT+C_4U)\frac{W}{a}+C_5\frac{1}{a}.
\end{equation}
Note that all the remaining terms in this equation are inverse proportional to
the angular diameter $a$.

Using the data about radial velocities, \mbox{H\,{\sc i}} line widths, angular diameters,
morphological types, surface brightness indices and axial ratio of galaxies it
is possible to obtain the value and errors for all $29$ parameters of the
relativistic model. For the previous version of the sample it was done by
\citet{ref:ParGayd05}. The statistical weights of all galaxies are assumed to
be the same. Besides the whole sample we also use subsamples limited in depth
$R<R_{max}$. They are defined as follows: they contain all galaxies, which have
Hubble velocity less than $R_{max}$ in the non-relativistic D-model. The
results of processing of the subsamples with $R_{max}=8000\,\rmn{km\,s}^{-1}$ and
$R_{max}=10000\,\rmn{km\,s}^{-1}$ are presented in Table \ref{tbl:1}. It contains
information about the standard deviation $\sigma$, coefficients of the Tully-Fisher
relation, apex coordinates and modulus of the bulk flow and parameters of the
multipole components. The results of processing of the same subsamples in the
non-relativistic model are given in the paper \citep{ref:APSS09}.

\begin{table*}
\begin{minipage}{155mm}
\caption{Parameters of the relativistic (R), semirelativistic (SR) and
semirelativistic with fixed $\gamma$ (SR$\gamma$) models of collective motion
for the subsamples with $R_{max}=10000\,\rmn{km\,s}^{-1}$ and $R_{max}=8000\,\rmn{km\,s}^{-1}$}
\begin{tabular}{lr@{${}\pm{}$}lr@{${}\pm{}$}lr@{${}\pm{}$}lr@{${}\pm{}$}lr@{${}\pm{}$}lr@{${}\pm{}$}l}
\hline
&\multicolumn{6}{c}{$R_{max}=10000\,\rmn{km\,s}^{-1}$}&\multicolumn{6}{c}{$R_{max}=8000\,\rmn{km\,s}^{-1}$}\\
&\multicolumn{2}{c}{R}&\multicolumn{2}{c}{SR}&\multicolumn{2}{c}{SR$\gamma$}&\multicolumn{2}{c}{R}&\multicolumn{2}{c}{SR}&\multicolumn{2}{c}{SR$\gamma$}\\
\hline
$\sigma,\,\rmn{km\,s}^{-1}$&\multicolumn{2}{c}{1130}&\multicolumn{2}{c}{1134}&\multicolumn{2}{c}{1157}&\multicolumn{2}{c}{1008}&\multicolumn{2}{c}{1018}&\multicolumn{2}{c}{1033}\\
$C_1$&$-0.35$&$0.11$&$-0.38$&$0.11$&$-0.39$&$0.11$&$-0.62$&$0.12$&$-0.60$&$0.12$&$-0.54$&$0.13$\\
$C_2$&$15.56$&$1.26$&$15.95$&$1.25$&$12.40$&$1.18$&$16.89$&$1.43$&$16.35$&$1.42$&$12.73$&$1.31$\\
$C_3$&$ 1.96$&$0.19$&$ 2.06$&$0.19$&$ 1.61$&$0.19$&$ 2.55$&$0.22$&$ 2.59$&$0.22$&$ 2.04$&$0.21$\\
$C_4$&$ 9.23$&$1.16$&$ 9.45$&$1.17$&$ 8.06$&$1.17$&$ 8.84$&$1.30$&$ 9.73$&$1.31$&$ 8.45$&$1.32$\\
$C_5$&$-1007$&$ 103$&$-1062$&$ 103$&$ -566$&$  80$&$-1095$&$ 106$&$-1120$&$ 107$&$ -683$&$  81$\\
$\gamma,\,10^{-6}\,\rmn{s\,km}^{-1}$&$-14.09$&$2.90$&$-16.62$&$2.55$&\multicolumn{2}{c}{$3.98$}&$-18.31$&$3.85$&$-19.59$&$3.59$&\multicolumn{2}{c}{$3.98$}\\
$D_x,\,\rmn{km\,s}^{-1}$&$ 164.8$&$93.8$&$ 111.2$&$91.7$&$   2.0$&$94.3$&$ 154.8$&$99.3$&$ 125.0$&$97.5$&$  42.7$&$100.3$\\
$D_y,\,\rmn{km\,s}^{-1}$&$-117.4$&$95.2$&$-191.8$&$90.1$&$-108.6$&$93.2$&$ -67.8$&$95.0$&$-100.2$&$93.4$&$-113.0$&$ 96.6$\\
$D_z,\,\rmn{km\,s}^{-1}$&$ 133.8$&$76.5$&$  64.6$&$72.0$&$ 144.4$&$73.9$&$  23.0$&$76.6$&$ -44.4$&$74.8$&$  32.4$&$ 76.1$\\
$D,\,\rmn{km\,s}^{-1}$&$243$&$79$&$231$&$80$&$181$&$74$&$171$&$90$&$166$&$89$&$125$&$87$\\
$l,b,\,deg$&\multicolumn{2}{c}{$325,\,34$}&\multicolumn{2}{c}{$300,\,16$}&\multicolumn{2}{c}{$271,\,53$}&\multicolumn{2}{c}{$336,\,8$}&\multicolumn{2}{c}{$321,\,-16$}&\multicolumn{2}{c}{$291,\,15$}\\
$q_1,\,10^{-2}$&$  4.90$&$4.28$&$ 6.84$&$1.39$&$ 6.68$&$1.65$&$ -3.33$&$5.11$&$  6.90$&$1.41$&$  7.04$&$ 1.67$\\
$q_2,\,10^{-2}$&$  5.76$&$4.66$&$-1.99$&$1.43$&$-3.98$&$1.70$&$  9.09$&$5.76$&$ -1.65$&$1.49$&$ -1.89$&$ 1.79$\\
$q_3,\,10^{-2}$&$ -3.11$&$5.85$&$-1.08$&$1.82$&$-0.47$&$2.17$&$ -1.49$&$7.39$&$ -1.10$&$2.02$&$ -0.31$&$ 2.41$\\
$q_4,\,10^{-2}$&$-17.36$&$6.47$&$ 2.22$&$2.20$&$ 3.79$&$2.62$&$-16.10$&$7.78$&$  0.99$&$2.17$&$  2.21$&$ 2.58$\\
$q_5,\,10^{-2}$&$ -9.54$&$7.53$&$ 0.37$&$2.40$&$-0.65$&$2.86$&$-35.03$&$9.08$&$  0.36$&$2.41$&$ -0.47$&$ 2.87$\\
$o_1,\,10^{-6}\,\rmn{s\,km}^{-1}$&$  3.18$&$1.21$&$ 2.35$&$1.14$&$ 3.40$&$1.62$&$  3.79$&$1.62$&$  3.34$&$1.58$&$  4.55$&$ 2.23$\\
$o_2,\,10^{-6}\,\rmn{s\,km}^{-1}$&$ -0.55$&$1.52$&$ 1.35$&$1.32$&$ 2.01$&$1.87$&$ -4.05$&$1.69$&$ -3.39$&$1.61$&$ -5.08$&$ 2.26$\\
$o_3,\,10^{-6}\,\rmn{s\,km}^{-1}$&$  5.05$&$1.45$&$ 5.13$&$1.37$&$ 7.77$&$1.95$&$  5.69$&$1.85$&$  5.78$&$1.82$&$  7.32$&$ 2.57$\\
$o_4,\,10^{-6}\,\rmn{s\,km}^{-1}$&$ -4.31$&$1.84$&$-3.64$&$1.70$&$-4.82$&$2.41$&$ -3.75$&$2.12$&$ -3.40$&$2.09$&$ -5.46$&$ 2.96$\\
$o_5,\,10^{-6}\,\rmn{s\,km}^{-1}$&$  3.58$&$1.60$&$ 3.71$&$1.49$&$ 4.39$&$2.12$&$  3.44$&$1.91$&$  3.82$&$1.88$&$  4.28$&$ 2.65$\\
$o_6,\,10^{-6}\,\rmn{s\,km}^{-1}$&$ -0.72$&$1.69$&$-1.86$&$1.56$&$-0.32$&$2.21$&$ -3.20$&$2.04$&$ -3.56$&$1.98$&$ -3.60$&$ 2.80$\\
$o_7,\,10^{-6}\,\rmn{s\,km}^{-1}$&$ 19.13$&$5.81$&$18.16$&$5.46$&$21.03$&$7.73$&$ 20.30$&$7.59$&$ 24.16$&$7.44$&$ 31.06$&$10.47$\\
$p_1,\,10^{-6}\,\rmn{s\,km}^{-1}$&$ -2.28$&$1.74$&$ 0.29$&$1.48$&$-2.49$&$2.09$&$  2.29$&$2.13$&$  4.66$&$2.01$&$  3.11$&$ 2.82$\\
$p_2,\,10^{-6}\,\rmn{s\,km}^{-1}$&$ -0.74$&$1.94$&$ 0.75$&$1.79$&$ 3.99$&$2.57$&$ -0.82$&$2.65$&$  0.73$&$2.55$&$  4.12$&$ 3.67$\\
$p_3,\,10^{-6}\,\rmn{s\,km}^{-1}$&$ -1.75$&$2.70$&$ 1.41$&$2.22$&$-5.15$&$3.06$&$ -4.28$&$2.80$&$ -2.69$&$2.65$&$ -5.95$&$ 3.80$\\
$s_1,\,10^{-6}\,\rmn{s\,km}^{-1}$&$  1.98$&$5.55$&\multicolumn{2}{c}{---}&\multicolumn{2}{c}{---}&$ 14.97$&$7.41$&\multicolumn{2}{c}{---}&\multicolumn{2}{c}{---}\\
$s_2,\,10^{-6}\,\rmn{s\,km}^{-1}$&$-10.88$&$5.82$&\multicolumn{2}{c}{---}&\multicolumn{2}{c}{---}&$-16.09$&$8.28$&\multicolumn{2}{c}{---}&\multicolumn{2}{c}{---}\\
$s_3,\,10^{-6}\,\rmn{s\,km}^{-1}$&$  2.84$&$7.05$&\multicolumn{2}{c}{---}&\multicolumn{2}{c}{---}&$  1.29$&$1.08$&\multicolumn{2}{c}{---}&\multicolumn{2}{c}{---}\\
$s_4,\,10^{-6}\,\rmn{s\,km}^{-1}$&$ 28.01$&$8.53$&\multicolumn{2}{c}{---}&\multicolumn{2}{c}{---}&$ 26.61$&$1.15$&\multicolumn{2}{c}{---}&\multicolumn{2}{c}{---}\\
$s_5,\,10^{-6}\,\rmn{s\,km}^{-1}$&$ 13.94$&$9.80$&\multicolumn{2}{c}{---}&\multicolumn{2}{c}{---}&$ 54.83$&$1.35$&\multicolumn{2}{c}{---}&\multicolumn{2}{c}{---}\\
\hline
\end{tabular}\label{tbl:1}
\end{minipage}
\end{table*}

\subsection{Semirelativistic model of collective motion}

The obtained results appear to have problems -- the values of $\tens{S}$ are
200 times larger then their estimation from equation (\ref{eqn:Weyl}). The same
situation appeared when processing the previous sample. It is easy to see that
the right-hand part of equation (\ref{eqn:Weyl}) is dominated by the term
containing tensor $\tens{S}$ due to the speed of light.

For the homogeneous isotropic Universe the Weyl tensor and tensors $\tens{S}$ and
$\tens{Q}$ vanish. They are connected with spatial inhomogeneities of density
distribution, e.g.  attractors and voids. The left part of this relation is a
sum of inputs of individual inhomogeneities. For a spherically symmetric
attractor with an excessive mass $M$ at a distance $u$ the spatial part of the
tensor $C_{\alpha\beta\gamma\delta}V^\beta V^\delta$ after reduction to
eigenaxes receives the form:
\begin{equation}
C_{\alpha\beta\gamma\delta}V^\beta V^\delta
=\frac{GM}{c^2u^3}\left(\begin{array}{ccc}2&0&0\\0&-1&0\\0&0&-1\end{array}\right).
\end{equation}

This value falls cubically with distance, so the main input is provided by
nearby attractors. In the paper \citep{ref:ParGayd05} the input of the Great
Attractor, Perseum-Pisces superclaster, Shepley concentration and Virgo cluster
were analyzed. The excessive masses and distances to attractors were taken from
the paper \citep{ref:Marinoni}. It was shown that the greatest input is
provided by the Virgo cluster. Adding the inputs of all attractors we obtain an estimation
of tensor $\tens{S}$, which appears to be 200 times smaller than the calculated
values.  This is caused by the same distribution of the tensors $\tens{Q}$ and
$\tens{S}$ over the celestial sphere. Due to measurement errors and deviations
from the Tully-Fisher relation the tensor $\tens{S}$ ``borrows'' some of the
value of tensor $\tens{Q}$. Unfortunately, with the quality and quantity of available
observational data we are unable to correctly separate the inputs of these tensors.
Thus, taking into account the small values of tensor $\tens{S}$, a so-called semirelativistic
model was introduced in the paper \citep{ref:ParGayd05}, which differs from the
full relativistic model only by the dropped term with tensor $\tens{S}$. So, it
will be possible to use the relativistic model only when we get samples with
significantly better quality and larger depth. For this reason, we switch to
the semirelativistic model (\ref{eqn:D}, \ref{eqn:Q}, \ref{eqn:O}, \ref{eqn:DQOR},
\ref{eqn:TFR}). The results of processing in semirelativistic model are also given in
Table \ref{tbl:1}. The semirelativistic DQ- and D-models are exactly the same as
their relativistic counterparts.

However, in both the relativistic and semirelativistic models there is a serious
problem. The calculated value of $\gamma$ appears to be way off the expected
value, namely $(-14.1\pm 2.9)\cdot 10^{-6}\,\rmn{s\,km}^{-1}$ in the relativistic
model and $(-16.6\pm 2.6) \cdot 10^{-6}\,\rmn{s\,km}^{-1}$ in the semirelativistic
model for $R_{max}=10000\,\rmn{km\,s}^{-1}$. This value essentially differs from
$\gamma_0=3.98\cdot 10^{-6}\,\rmn{s\,km}^{-1}$ (\ref{eqn:g0}) calculated from
cosmological parameters. As one can see from Table \ref{tbl:1}, the second term in
equation (\ref{eqn:gR}) is negligible and cannot be responsible for the discussed
effect. Naturally, we do not question the values of the cosmological parameters
and the reason should be sought elsewhere.

In the next two section we consider the reasons, which could lead to the
deviation of the calculated value of $\gamma$ from its value (\ref{eqn:g0})
obtained from the cosmological parameters. These reasons can be caused either
by the dependence of linear galaxy diameters on distance or by the influence of
measurement errors.

\section{Shift of \mbox{\boldmath{$\gamma$}} due to dependence of linear galaxy diameters on distance}

When determining the distances to the galaxies using the generalised Tully-Fisher
relation in the ``linear diameter -- \mbox{H\,{\sc i}} line width'' version, we assume that
galaxies with the same \mbox{H\,{\sc i}} line width, morphological type, axial ratio and
surface brightness index have the same linear diameter $L$. Let us consider the opposite
case when the linear diameter weakly depends on the distance $r$ according to the law
\begin{equation}
L(r)=L_0+r\frac{dL}{dr}.
\end{equation}
Here $L_0$ is the linear diameter of nearby galaxies according to Tully-Fisher
relation. On the other hand, if we express the angular diameters $a$ in
radians, the linear diameters will be given by
\begin{equation}
L(r)=ar.
\end{equation}
This yields a problem: we determine the distances to the galaxies using the
Tully-Fisher relation in an assumption that the linear diameters of the
galaxies are equal to $L_0$ rather than $L(r)$. Thus, the apparent distance
$\rho$ will be equal to
\begin{equation}
\rho=\frac{L_0}{a}.
\end{equation}
Combining the latter two formulae we can express the true distance $r$ through
the apparent distance $\rho$:
\begin{equation}
r=\rho\left(1+\frac{r}{L_0}\frac{dL}{dr}\right).
\end{equation}

Let us calculate the shift of $\gamma$ due to this effect. Radial velocities of
galaxies according to equation (\ref{eqn:R}) are given by
\begin{equation}
V=Hr+\gamma_0(Hr)^2,
\end{equation}
where $\gamma_0$ is the true value of $\gamma$. However, instead of this
formula we use the following expression:
\begin{equation}
V=H\rho+\gamma(H\rho)^2.
\end{equation}
It is trivial to find that
\begin{equation}\label{eqn:dg}
\Delta\gamma=\gamma-\gamma_0=\frac{1}{HL_0}\frac{dL}{dr}.
\end{equation}

There are several effects leading to the dependence (\ref{eqn:dg}). Let us
consider them one by one.

\subsection{Effect of galaxy evolution}

Let us assume that galaxies evolve with time changing their linear diameters
with a characteristic rate $\dot{L}=dL/dt$. Since we observe more
distant galaxies at earlier stages of development, we can write
\begin{equation}\label{eqn:evol}
\frac{dL}{dr}=-\frac{\dot{L}}{c},\,\Delta\gamma=-\frac{1}{Hc}\frac{\dot{L}}{L}.
\end{equation}
This effect is described, in particular, by \citet{ref:Weinberg}.

The question is, whether it alone can explain the observed value of
$\Delta\gamma\sim 2.2\cdot 10^{-5}\,\rmn{s\,km}^{-1}$? A simple estimation shows that
would this be the case, the galaxies should shrink with a typical rate of
$4.7\cdot 10^{-10}\,\rmn{yr}^{-1}$. This value contradicts to our knowledge of galaxy
evolution. In particular, such large shrinking rate would lead to decrease of
galaxies by about 12 per cent in one orbital period of the Sun in the Milky
Way. Thus, this effect cannot make an essential contribution to the observed
shift of $\gamma$.

\subsection{Effect of selection by angular diameters}\label{s:sel}

The Tully-Fisher relation is statistical. In fact, galaxies can be smaller or
larger than given by it. For galaxies with angular diameters close to the
threshold $a_b=0\farcm 6$ there is a selection. Large galaxies will enter RFGC
and the small ones will have too small angular diameters and will be rejected.
Thus, the average linear diameters of RFGC galaxies will increase with distance.
This is usually referred to as Malmquist bias. According to equation
(\ref{eqn:dg}), these effects will lead to a positive shift of $\Delta\gamma$.
This shift has the opposite sign to the observed one and, therefore, cannot be
its cause. Nevertheless, this effect cannot be totally neglected and further we
will estimate its value using Monte Carlo simulations.

\subsection{Effect of cosmological decrease of observed surface brightness of galaxies}

Consideration of space-time curvature even in the simplest homogeneous
isotropic cosmological models leads to a number of effects. One of these
effects yields the decrease of observed surface brightness of galaxies. It is
quite evident that the surface brightness is proportional to $(r/D)^2$, where D
is the photometric distance. In flat space-time there is no difference between
$r$ and $D$ and the surface brightness does not depend on the distance. In
curved space-time at small $z$ we can use the expressions for $r$ and $D$ from
the book \citep{ref:ZeldNov}, which yield $I=I_{NR}(1-4z)$. Here $I_{NR}$ is
the surface brightness in the flat space-time and $z$ is the redshift. This
expression is a low-z limit of the Tolman effect \citep{ref:Tolman30,ref:Tolman34},
which describes the decrease of surface brightness by a factor of $(1+z)^4$. Switching to
the brightness $\mu=-2.5 \mathrm{lg} I$ measured in $\rmn{mag}/\sq\arcsec$ we
obtain the shift
\begin{equation}\label{eqn:dmu}
\begin{array}{l}
\Delta\mu =\mu-\mu_{NR}=\frac{10}{\mathrm{ln} 10}R/c\\
\phantom{\Delta\mu =}{}=1.45\cdot 10^{-5}R\,\rmn{mag}/\sq\arcsec\,\rmn{s\,km}^{-1}.
\end{array}
\end{equation}
For a galaxy at the distance $100 h^{-1}\,\rmn{Mpc}$ this gives
$\Delta\mu=0.145\,\rmn{mag}/\sq\arcsec$. Earlier such effects were considered,
e.g. by \citet{ref:Sandage}.

If galaxies had sharp edges where the surface brightness instantly
vanishes, this effect would be of no interest for us. However, for real
galaxies the surface brightness gradually falls to the edges according to the
de Vaucouleurs law $I(l)=I_0e^{-l/l_0}, \mu(l)=\mu_0+1.0857\,l/l_0$
\citep{ref:Vac59}. Here $I_0$ and $\mu_0$ are the values of $I$ and $\mu$ in
the centre of the galaxy, $l$ is the radial distance from the centre of galaxy
and $l_0$ is a characteristic radial scale. For spiral galaxies the scale $l_0$ weakly
depends on the morphological type. For bright, comparable to the Milky Way,
galaxies it ranges from 1 to 10 kpc \citep{ref:Resh}. A more accurate estimation
was obtained by \citet{ref:Fathi10} using the sample of more than
30000 galaxies. The average value of $l_0$ appeared to be equal to $3.8\pm
2.1\,\rmn{kpc}$. For smaller galaxies with the total stellar mass
$10^9-10^{10}\,M_{\sun}$ they estimated it as $1.5\pm 0.7\,\rmn{kpc}$ and for
larger galaxies with the total stellar mass $10^{11}-10^{12}\,M_{\sun}$ -- as
$5.7\pm 1.9\,\rmn{kpc}$. For the later types of galaxies, which are predominant
in RFGC, the value of $l_0$ does not typically exceed $2\,\rmn{kpc}$, according to
\citet{ref:Freeman70}. In his sample of 36 galaxies the maximum value of $l_0$
was $6.1\,\rmn{kpc}$, and the second largest was $4.5\,\rmn{kpc}$.

The linear diameters of RFGC galaxies are determined at the isophotal level
$\mu=25\,\rmn{mag}/\sq\arcsec$. Due to the abovementioned effect for distant galaxies
this boundary is shifted with respect to its position in flat space-time. For
instance, for a galaxy at $r=100 h^{-1}\,\rmn{Mpc}$ the region corresponding to the
isophotal level $25\,\rmn{mag}/\sq\arcsec$ in the flat space-time would appear
at the isophotal level $25.145\,\rmn{mag}/\sq\arcsec$ in the real Universe.

For this reason, the isophotal boundaries will be shifted towards the centre of
the galaxy by $\delta l=1.33\cdot10^{-5}l_0R\,\rmn{s\,km}^{-1}$. Naturally, the
apparent decrease of isophotal diameter $\delta L$ will be twice this value.
Using this distance-dependent isophotal diameter instead of constant linear
diameter, we obtain the so-called isophotal distance instead of the angular
diameter distance.

The expression (\ref{eqn:g}) is derived for angular distance only. In reality,
however, we deal with isophotal distances with a slightly different value of
$\gamma$. From the equations (\ref{eqn:dmu}) and $L(r)=L_0-\delta L$ we get
$\delta\gamma = -2.66\cdot 10^{-5} l_0/L_0\,\rmn{s\,km}^{-1}$. Let us estimate this
value. As a typical galaxy diameter we take the value for the Milky Way:
$L_0=30\,\rmn{kpc}$. The distribution of $\mathrm{lg} L_0$ for RFGC galaxies, where
$L_0$ is expressed in kpc, is given by \citet{ref:Kud97}. The maximum of this
distribution corresponds to the interval from 1.3 to 1.4, ehich corresponds to
$L_0\sim 22\,\rmn{kpc}$.
The mean value of $L_0$ for RFGC galaxies should be taken slightly larger because
$\left<L_0\right>>10^{\left<\mathrm{lg} L_0\right>}$. With $L_0=30\,\rmn{kpc}$
and $l_0=3\div 5\,\rmn{kpc}$ we get $\Delta\gamma=(-2.7\div -4.4)\cdot 10^{-6}\,\rmn{s\,km}^{-1}$,
which constitutes from 12 to 20 per cent of observed shift. Even the extreme
estimation with $L_0=22\,\rmn{kpc}, l_0=10\,\rmn{kpc}$ can explain only 50 per cent of the
observed shift. For edge-on spiral galaxies the surface photometry was
performed in a series of articles by van der Kruit and Searle. The data are
assembled in Table 3 of the paper \citep{ref:vdKS3}. For 7 galaxies, 4 of which
enter the RFGC (NGC4244=RFGC2245; NGC5907=RFGC2946; NGC4565=RFGC2335; NGC5023=RFGC2495),
the ratio $l_0/L_0$ ranges from 0.10 to 0.15, which corresponds to
$\Delta\gamma=(-4\div -6)\cdot 10^{-6}\,\rmn{s\,km}^{-1}$.

This effect can be responsible for the observed shift, but only partially.
Thus, we still need to find the reason behind the main share of the shift.

\section{Shift of \mbox{\boldmath{$\gamma$}} due to the influence of measurement errors}

Let us show that the observed shift can be explained with purely statistical
effects due to measurement errors of \mbox{H\,{\sc i}} line widths and angular diameters.

\subsection{Estimating the impact of measurement errors: a simple case}\label{s:MB1}

Before trying to address this problem at its full extent, let us consider a
simple case when an analytical solution can be provided. Let us start off from
introducing a toy model $y=Ax+Bx^2$, where $x=W/a$ is the main term of the
Tully-Fisher relation (\ref{eqn:TFR}), $y=V$, $A=C_1$, and $B=\gamma_0 C_1^2$.
For generality we consider not only the value of $\gamma_0$ given by equation
(\ref{eqn:g0}) but any fixed value. This model corresponds to an isotropic
Hubble expansion with cosmological acceleration.

Observational data provide us with a set of $N$ points characterized by values
$x_i$ and $y_i$. It is important to realize how measurement errors and
deviations from the Tully-Fisher relation distort the dataset. The errors in
velocity measurements and deviations from Tully-Fisher relation yield errors in
$y$. The values of $A$ and $B$ determined by the least square method from the
dataset with such errors will have normal distribution of errors without shift.
At the same time, errors in measurements of $W$ or $a$ yield errors in $x$.
This case is similar to Malmquist bias. Due to this effect, the perceived values
of $A$ and $B$ given by the least square method will have a systematic error.

\begin{figure}
\begin{minipage}{84mm}
\includegraphics[width=84mm]{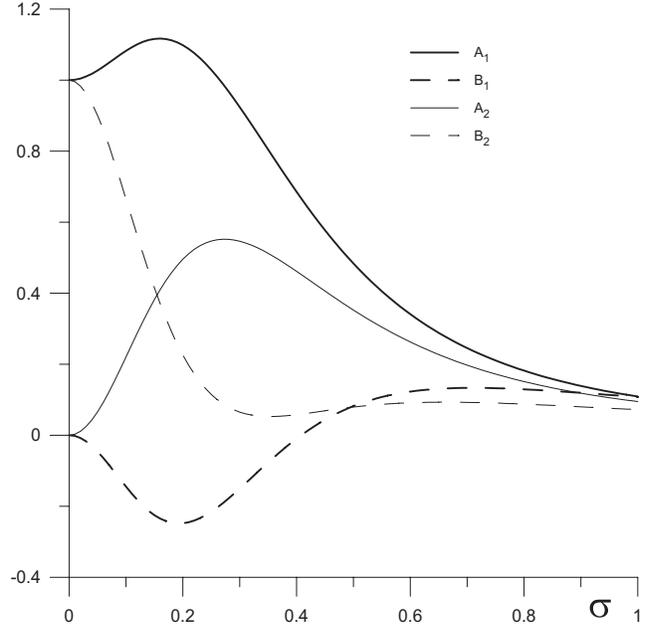}
\caption{Dependence of the coefficients of equations (\ref{eqn:a}) and (\ref{eqn:b})
on the noise level}
\label{fig:1}
\end{minipage}
\end{figure}

Let us consider the following case: we have $N$ values of $x_i$ distributed
uniformly over the interval $[0,1]$ with a step $(N-1)^{-1}$. The values of $y$
are calculated with $A=A_0=1$ and $B=B_0=\gamma_0$. In each of $N$ points the $x_i$ is
shifted by $\sigma \xi_i$, where $\xi_i$ is a normally distributed quantity
with zero mean and unit variance. The different values of $\xi_i$ are not
correlated with each other. The values $y_i$ are calculated from the original
nondisplaced values of $x_i$. The values of $A$ and $B$ given by the least
square method have the form
\begin{equation}
A=\frac{\sum\limits_{i=1}^N{y_ix_i}\sum\limits_{i=1}^N{x_i^4}-\sum\limits_{i=1}^N{y_ix_i^2}\sum\limits_{i=1}^N{x_i^3}}
{\sum\limits_{i=1}^N{x_i^2}\sum\limits_{i=1}^N{x_i^4}-\left(\sum\limits_{i=1}^N{x_i^3}\right)^2},
\end{equation}
\begin{equation}
B=\frac{\sum\limits_{i=1}^N{x_i^2y_i}\sum\limits_{i=1}^N{x_i^2}-\sum\limits_{i=1}^N{x_i^3}\sum\limits_{i=1}^N{x_iy_i}}
{\sum\limits_{i=1}^N{x_i^2}\sum\limits_{i=1}^N{x_i^4}-\left(\sum\limits_{i=1}^N{x_i^3}\right)^2},
\end{equation}
\begin{equation}
x^0_i=\frac{i-1}{N-1},\,x_i=x^0_i+\sigma \xi_i,\,y_i=x^0_i+\gamma_0(x^0_i)^2.
\end{equation}
It is not very difficult to calculate the mean values of $A$ and $B$ over $\xi$
using the following expressions:
\begin{equation}\label{eqn:a}
A=A_1+\gamma_0A_2,
\end{equation}
\begin{equation}\label{eqn:b}
B=B_1+\gamma_0B_2.
\end{equation}
Here we designated
\begin{equation}\label{eqn:a1}
A_1=\frac{1+40\sigma^2+60\sigma^4}{1+28\sigma^2+180\sigma^4+720\sigma^6},
\end{equation}
\begin{equation}\label{eqn:a2}
A_2=\frac{28\sigma^2+60\sigma^4}{1+28\sigma^2+180\sigma^4+720\sigma^6},
\end{equation}
\begin{equation}\label{eqn:b1}
B_1=\frac{-20\sigma^2+120\sigma^4}{1+28\sigma^2+180\sigma^4+720\sigma^6},
\end{equation}
\begin{equation}\label{eqn:b2}
B_2=\frac{1-43/3\sigma^2+80\sigma^4}{1+28\sigma^2+180\sigma^4+720\sigma^6}.
\end{equation}
Note that the formulae (\ref{eqn:a1}, \ref{eqn:a2}, \ref{eqn:b1}, \ref{eqn:b2})
are precise up to $\mathcal{O}(1/N)$. To verify
these formulae we calculated $A$ and $B$ for this toy model using $10000$ Monte
Carlo simulations. The mean values perfectly fitted the given formulae (see
Figure \ref{fig:1}).

\begin{figure}
\begin{minipage}{84mm}
\includegraphics[width=84mm]{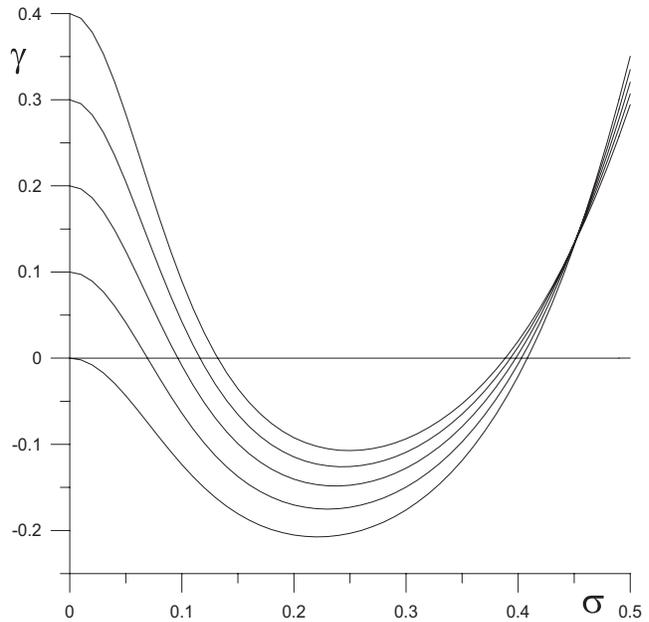}
\caption{Dependence of least squares estimation of $\gamma$ on the noise level
for different values of $\gamma_0$. The values of $\gamma_0$ equal to values of
$\gamma$ at $\sigma=0$}
\label{fig:2}
\end{minipage}
\end{figure}

If we calculate $\gamma$ using the formula $\gamma=B/A^2$, we will obtain
instead of true value $\gamma_0$ a value $\gamma$, plotted on Figure
\ref{fig:2} against $\sigma$ for different $\gamma_0$. At $\sigma=0$, i.e. when
there are no errors, we obtain $\gamma=\gamma_0$, but at small $\sigma$ we
obtain $\gamma<\gamma_0$. This is the impact of the measurement errors we
demonstrate.

\subsection{Estimating the impact of measurement errors: a real case}\label{s:MB2}

The actual case is much more complicated. The problem can not be reduced to the
one-dimensional case, since the galaxy's velocity depends on its position on
the celestial sphere. To estimate the distance we use all the terms in the
relation (\ref{eqn:TFR}). The errors in determination of distances are
non-Gaussian. They are due to the errors in angular diameters and \mbox{H\,{\sc i}} line widths
and deviations from the Tully-Fisher relation. These errors were analyzed in
the paper \citep{ref:ParPar08}. Here we will briefly mention the main points of
the routine used. These errors can be described by four parameters:
\begin{equation}\label{eqn:vrel}
V_i=V^{(0)}_i(1+s_V\xi_1),
\end{equation}
\begin{equation}\label{eqn:wrel}
W_i=W^{(0)}_i(1+s_W\xi_2),
\end{equation}
\begin{equation}\label{eqn:arel_aabs}
a_i=a^{(0)}_i(1+s_a\xi_3)+\Delta_a\xi_4.
\end{equation}
Here $\xi_1$, $\xi_2$, $\xi_3$, and $\xi_4$ are the four independent
non-correlating random values. They are distributed according to Gauss law with
zero mean and unit variance.

Let us describe what errors correspond to each type of noise. The noise
(\ref{eqn:vrel}) describes the deviations from the Tully-Fisher relationship.
It does not include velocity measurement error, because radial velocities are
well determined, as well as the directions towards galaxies. Such a type of
noise provides a conventional log-normal distribution of the velocity
deviation.  If this noise is used alone, the maximal value of $s_V$ can be
assumed about $0.2$, which corresponds to $20$ per cent uncertainty of distance
estimated by the Tully-Fisher relation.

The noise (\ref{eqn:wrel}) corresponds to \mbox{H\,{\sc i}} line width measurement errors.
Different methods of calculating \mbox{H\,{\sc i}} half-width of the same profile can differ
by up to $10$ per cent. Thus, if this noise is used alone, the maximal value of
$s_W$ can be set to $0.15$ with some tolerance.

The noise (\ref{eqn:arel_aabs}) describes angular diameter measurement errors.
The value $s_a$ describes a relative error, caused by variations of exposition,
curvature of galaxies etc. The diameter measurements have an error of about
$5\div 10$ per cent (Karachentseva, private communication). The value
$\Delta_a$ describes an absolute error of measurement. This value is important
for the smallest galaxies. Since these diameters were measured in tenths of
millimetre, which corresponded to 0.11 arcmin, the value of $\Delta_a$
can be estimated between 0.05 and 0.1. Note that if we use non-zero values of
$s_a$ and $\Delta_a$ simultaneously, the maximum estimation of $\Delta_a$
should be somewhat reduced to avoid overestimating the noise added to angular
diameters for the smallest galaxies.

Naturally, such a difficult problem of determining the shift of $\gamma$ can
not be solved analytically. We use Monte Carlo simulations to resolve it
numerically. Let us describe the details of this procedure.

At first, we use a subsample with $R_{max}=10000\,\rmn{km\,s}^{-1}$, which contains
$N=1459$ galaxies. For each of the galaxies we substitute the measured radial
velocity with the radial velocity calculated using the formulae
(\ref{eqn:D}, \ref{eqn:Q}, \ref{eqn:O}, \ref{eqn:DQOR}, \ref{eqn:TFR}). We use
the values given by equations (\ref{eqn:g}) and (\ref{eqn:q}).

After that we add noise to our model and for each of $10000$ simulation we
calculate the coefficients in the same way as we treat actual data, i.e. using
the semirelativistic model. Thus, for each realisation we obtain a complete set
of the coefficients including $\gamma$. For the obtained values of $\gamma$ we
calculate the mean and the standard deviation. Taking into account that the
distribution of $\gamma$ is non-Gaussian, its quantiles differ from the ones
calculated from the normal distribution. In this article we give errors
corresponding to the $99$ per cent confidence level. Application of Monte Carlo
method allows to do this in a straightforward way. From $10000$ values of
$\gamma$ for different realizations we find the 50th largest and smallest
values. They give us the boundaries of the $99$ per cent confidence interval.

In principle, we can apply this procedure to any mock catalogue. However, we
should use a catalogue that has the same spatial distribution as well as
distribution of morphological types, surface brightness index etc. as the main
sample. The best mock catalogue is thus the sample itself. We use as a result
the sample of actually measured parameters, namely angular diameter, surface
brightness index, Hubble type, \mbox{H\,{\sc i}} line width, ratio of angular diameters in
red and blue imprints, and celestial coordinates. The radial velocity is
calculated from the formulae (\ref{eqn:R}, \ref{eqn:TFR}) with coefficients
obtained from the real sample. Hwever, we use a more refined procedure
to improve reliability.

At first we calculated the coefficients for the semirelativistic model using
real data while fixing the value of $\gamma$ at $\gamma_0$ (\ref{eqn:g0}). The
corresponding coefficients are also given in Table \ref{tbl:1}. Then we used
this set of coefficients to calculate mock radial velocities for the galaxies.
Such mock radial velocities are closer to the actual radial velocities than for
any other models with fixed $\gamma$.

The values obtained using the Monte Carlo method for different parameters of
the noise are given in Table \ref{tbl:2}. The top part of Table \ref{tbl:2}
illustrates the impact of each individual type of noise. One can see that the
$\gamma$ value acts in the same way as in the simple case considered in the
previous section. The noise $s_V$, corresponding to the deviations from the
statistical Tully-Fisher relation has little or no effect on the value of
$\gamma$. All other types of noise lead to a drastic reduction of $\gamma$,
especially the noise $\Delta_a$. The bottom part of Table \ref{tbl:2} contains
the results obtained with realistic noise parameters. When choosing parameters
we used as a control parameter the standard deviation $\sigma$ for the noised sample,
comparing it to the corresponding value for real data, given in Table \ref{tbl:1}. This
parameter is convenient because it vanishes in the absence of the noise and
grows when the noise increases. Thus, it can help avoiding undernoising and
overnoising. Other constraints used and details of the procedure are described
in the paper \citep{ref:ParPar08}. Naturally, we do not try to find a unique set
of noise parameters. The four noise values form a four-dimensional parameter
space, the $\sigma$ constraint yields a three-dimensional hypersurface in it.
Other constraints give us rough estimates of the boundaries of the volume of
suitable values.

\begin{table*}
\settowidth{\tmpla}{$8.8$}
\settowidth{\tmplb}{$-88.8$}
\settowidth{\tmplc}{~~~($-88.8; -88.8$)}
\begin{minipage}{155mm}
\caption{Results of Monte Carlo simulations, $3\sigma\,\rmn{CL}$ stands for 99 per cent confidence interval}
\begin{tabular}{ccccr@{${}\pm{}$}lr@{${}\pm{}$}l@{~~~(}r@{;~}r@{)~~~}r@{${}\pm{}$}lr@{${}\pm{}$}l@{~~~(}r@{;~}r@{)}}
\hline
$s_V$&$s_W$&$s_a$&$\Delta_a$&
\multicolumn{12}{@{}c@{}}{\begin{tabular}{@{}r@{${}\pm{}$}lr@{${}\pm{}$}l@{~~~(}r@{;~}r@{)~~~}r@{${}\pm{}$}lr@{${}\pm{}$}l@{~~~(}r@{;~}r@{)}}
\multicolumn{6}{@{}c@{}}{$R_{max}=10000\,\rmn{km\,s}^{-1}$}&
\multicolumn{6}{@{}c@{}}{$R_{max}= 8000\,\rmn{km\,s}^{-1}$}\\
\hline
\multicolumn{2}{@{}c@{}}{$\sigma,\,\rmn{km\,s}^{-1}$}&
\multicolumn{4}{@{}c}{\begin{tabular}{@{}r@{${}\pm{}$}l@{}c}
\multicolumn{3}{@{}c@{}}{$\gamma, 10^{-6}\,\rmn{s\,km}^{-1}$}\\
\hline
\hbox to 5mm {\hfil $\left<\gamma\right>$}&\hbox to \tmpla {\hfil $\sigma_{\gamma}$}&
\hbox to \tmplc {\hfil $3\sigma\,\rmn{CL}$~~~~~}\\
\end{tabular}}&
\multicolumn{2}{@{}c@{}}{$\sigma,\,\rmn{km\,s}^{-1}$}&
\multicolumn{4}{@{}c@{}}{\begin{tabular}{@{}r@{${}\pm{}$}l@{}c@{}}
\multicolumn{3}{@{}c@{}}{$\gamma, 10^{-6}\,\rmn{s\,km}^{-1}$}\\
\hline
\hbox to 5mm {\hfil $\left<\gamma\right>$}&\hbox to \tmpla {\hfil $\sigma_{\gamma}$}&
\hbox to \tmplc {\hfil $3\sigma\,\rmn{CL}$~~~~~}\\
\end{tabular}}\\
\end{tabular}}\\
\hline
\multicolumn{16}{c}{Without correction for selection}\\
\hline
$0.05$&$0.00$&$0.00$&$0.00$&$ 282$&$ 7$&$  4.0$&$1.2$&$  1.1$&$  7.2$&$ 245$&$ 6$&$  4.0$&$1.5$&$ -0.9$&$ 10.1$\\
$0.10$&$0.00$&$0.00$&$0.00$&$ 563$&$14$&$  4.1$&$2.4$&$ -1.8$&$ 10.4$&$ 491$&$13$&$  4.1$&$3.0$&$ -6.7$&$ 16.6$\\
$0.15$&$0.00$&$0.00$&$0.00$&$ 845$&$21$&$  4.2$&$3.6$&$ -4.1$&$ 14.3$&$ 736$&$19$&$  4.3$&$4.4$&$-12.8$&$ 22.1$\\
$0.20$&$0.00$&$0.00$&$0.00$&$1126$&$28$&$  4.4$&$4.8$&$ -6.7$&$ 18.5$&$ 981$&$25$&$  4.6$&$5.9$&$-15.8$&$ 30.8$\\
$0.00$&$0.05$&$0.00$&$0.00$&$ 312$&$ 7$&$ -0.6$&$1.1$&$ -3.4$&$  2.5$&$ 283$&$ 7$&$ -3.2$&$1.3$&$ -8.0$&$  2.4$\\
$0.00$&$0.10$&$0.00$&$0.00$&$ 592$&$14$&$ -9.4$&$1.5$&$-12.8$&$ -5.2$&$ 531$&$13$&$-15.7$&$1.6$&$-20.5$&$ -8.6$\\
$0.00$&$0.15$&$0.00$&$0.00$&$ 831$&$19$&$-16.5$&$1.3$&$-19.4$&$-12.8$&$ 737$&$17$&$-24.5$&$1.2$&$-28.2$&$-18.8$\\
$0.00$&$0.20$&$0.00$&$0.00$&$1031$&$23$&$-20.9$&$1.0$&$-23.3$&$-17.9$&$ 907$&$21$&$-29.4$&$0.9$&$-32.6$&$-25.2$\\
$0.00$&$0.00$&$0.05$&$0.00$&$ 281$&$ 7$&$ -2.2$&$1.0$&$ -4.8$&$  0.5$&$ 247$&$ 6$&$ -4.4$&$1.2$&$ -9.9$&$  0.0$\\
$0.00$&$0.00$&$0.10$&$0.00$&$ 538$&$13$&$-14.4$&$1.2$&$-17.2$&$-11.3$&$ 468$&$11$&$-19.8$&$1.3$&$-24.2$&$-14.2$\\
$0.00$&$0.00$&$0.15$&$0.00$&$ 756$&$17$&$-23.5$&$0.8$&$-25.4$&$-21.3$&$ 648$&$15$&$-30.2$&$0.8$&$-32.7$&$-25.8$\\
$0.00$&$0.00$&$0.20$&$0.00$&$ 946$&$23$&$-27.3$&$1.0$&$-29.1$&$-23.3$&$ 802$&$21$&$-33.8$&$1.0$&$-36.2$&$-21.6$\\
$0.00$&$0.00$&$0.00$&$0.05$&$ 370$&$11$&$ -8.2$&$1.4$&$-12.0$&$ -4.6$&$ 307$&$10$&$-10.4$&$1.6$&$-16.9$&$ -4.3$\\
$0.00$&$0.00$&$0.00$&$0.10$&$ 664$&$18$&$-21.9$&$1.2$&$-24.3$&$-16.7$&$ 548$&$15$&$-27.2$&$1.2$&$-31.3$&$ -1.0$\\
\hline
$0.17$&$0.08$&$0.03$&$0.06$&$1155$&$28$&$-17.0$&$2.2$&$-23.6$&$ -5.9$&$1004$&$25$&$-22.6$&$2.4$&$-30.5$&$-11.4$\\
$0.17$&$0.06$&$0.04$&$0.06$&$1125$&$27$&$-16.0$&$2.3$&$-23.3$&$ -3.9$&$ 976$&$25$&$-21.1$&$2.7$&$-28.8$&$ -9.1$\\
$0.17$&$0.07$&$0.04$&$0.06$&$1142$&$28$&$-16.8$&$2.2$&$-23.7$&$ -6.0$&$ 992$&$25$&$-22.2$&$2.6$&$-29.8$&$-11.2$\\
$0.18$&$0.04$&$0.02$&$0.08$&$1182$&$29$&$-18.8$&$2.1$&$-24.9$&$ -8.8$&$1020$&$26$&$-23.6$&$2.4$&$-31.5$&$-13.0$\\

$0.18$&$0.05$&$0.02$&$0.07$&$1169$&$29$&$-16.8$&$2.3$&$-23.5$&$ -5.4$&$1011$&$26$&$-21.5$&$2.7$&$-30.0$&$ -9.7$\\
$0.18$&$0.05$&$0.03$&$0.07$&$1174$&$29$&$-17.2$&$2.3$&$-24.8$&$ -6.3$&$1016$&$26$&$-22.0$&$2.7$&$-30.0$&$-10.3$\\
$0.18$&$0.05$&$0.04$&$0.07$&$1181$&$29$&$-17.7$&$2.2$&$-24.9$&$ -5.9$&$1022$&$26$&$-22.8$&$2.6$&$-30.1$&$-11.1$\\
\hline
\multicolumn{16}{c}{With correction for selection}\\
\hline
$0.18$&$0.05$&$0.02$&$0.07$&$1168$&$29$&$-15.6$&$2.4$&$-23.9$&$ -4.9$&$1009$&$26$&$-20.5$&$2.8$&$-29.0$&$-12.1$\\
$0.18$&$0.05$&$0.03$&$0.07$&$1173$&$29$&$-16.0$&$2.4$&$-24.1$&$ -5.2$&$1014$&$26$&$-21.0$&$2.7$&$-29.2$&$-13.0$\\
$0.18$&$0.05$&$0.04$&$0.07$&$1180$&$29$&$-16.6$&$2.3$&$-24.0$&$ -5.8$&$1020$&$26$&$-21.7$&$2.6$&$-30.5$&$-13.8$\\
\hline
\end{tabular}\label{tbl:2}
\end{minipage}
\end{table*}

One can see from Table \ref{tbl:2} that for a set of parameters lying inside or
close to those boundaries, the value of $\gamma$ is much less than the initial
value $\gamma_0$. Comparing these values with the value $\gamma=(-16.6\pm 2.6)
\cdot 10^{-6}\,\rmn{s\,km}^{-1}$ obtained from the real data, we can select a range
of suitable noise parameters. For many realistic noise parameters the value of
$\gamma$ falls into the $1\sigma$ area, and for most of them it falls into $99$
per cent confidence area. In those cases when $\gamma$ misses the $99$ per cent
confidence area, the value of $\Delta_a$ is unrealistically large.

We applied the same routine to the subsample with $R_{max}=8000\,\rmn{km\,s}^{-1}$.
The results for this subsample are also presented in Table \ref{tbl:2}. One can
see that there are sets of noise parameters, which provide suitable shift of
$\gamma$ for both subsamples. Nevertheless, we should mention that the same
parameters provide a much less suitable shift of $\gamma$ for the subsample
with $R_{max}=6000\,\rmn{km\,s}^{-1}$, where the observed value is $\gamma=(-11.0\pm
7.6) \cdot 10^{-6}\,\rmn{s\,km}^{-1}$. This is due to large errors in determination
of $\gamma$ for subsamples with low depth.

An additional advantage of this method is that by slight modification of the
algorithm we can also estimate the influence of selection described in section
\ref{s:sel}. For this purpose one should add an additional condition when
adding noise to angular diameters: if the noised angular diameter becomes less
than the minimal angular diameter in the actual sample, the noise should be
reapplied. This procedure reduces the shift of $\gamma$, but this effect is not
very strong, e.g. for $R_{max}=10000\,\rmn{km\,s}^{-1}$ it gives $\gamma=(-15.6\pm
2.4) \cdot 10^{-6}\,\rmn{s\,km}^{-1}$ against $\gamma=(-16.8\pm 2.3) \cdot
10^{-6}\,\rmn{s\,km}^{-1}$. For other subsamples this effect has the same order.
The best noise parameters for this case are given in the bottom part of Table \ref{tbl:2}.

\section{Errors in determination of distance caused by shift of \mbox{\boldmath{$\gamma$}}}

An important result of shift of $\gamma$ is that the distances calculated by
the relativistic Tully-Fisher relation (\ref{eqn:TFR}) become more than the
correct ones. Really, since the combination $R+\gamma R^2$ is fixed by redshift
data, the decrease of $\gamma$ leads to increase of $R$. We made some Monte
Carlo simulations which showed that this increase is about 18 per cent. Such
large errors will yield too large errors in determination of peculiar
velocities, defined in a standard manner $V_{pec}=V-R$. This is a great
drawback of relativistic and semirelativistic models. Note that after switching
to curved space-time, it is possible to introduce an alternative definition of
a peculiar velocity $V_{pec}=V-V_{cosm}$, where $V_{cosm}$ is the velocity of
cosmological expansion, which is defined in the low-$z$ limit as
$V_{cosm}=R+\gamma R^2$. Peculiar velocities defined in this way suffer much
less from this effect. Indeed, the non-relativistic model implicitly uses this
second definition of peculiar velocity. If we take a look at the generalized
Tully-Fisher relation (\ref{eqn:TF}) we will see that the term with $C_5$ is,
in fact, quadratic in distance. Thus, the 18 per cent difference in
distances leads to underestimation of the quadrupole components by a factor of
$1.18$, and of the octopole one -- by a factor of $(1.18)^2$. The dipole component remains
unaltered.

To correctly determine distances in relativistic or semirelativistic models we
have two options. The first one is to introduce a correction for this effect.
The second, technically easier, is to eliminate the origin of this effect by
fixing the value of $\gamma$ at $\gamma_0$. In this way we use the information
about cosmological constants obtained by other more precise methods. Thus we
naturally switch to the next stage -- the semirelativistic model with fixed
value of $\gamma$. Its 23 free parameters are given in Table \ref{tbl:1}. We
also performed Monte Carlo simulations with the same noise parameters in this
model. The difference between actual and estimated distances appeared to be
about 0.25 per cent. Naturally, the same procedure can be applied to
relativistic DQ- and D-models. In the next section we consider the collective
velocity field obtained in the framework of this model of galaxy motion.

\section{The multipole structure of the velocity field}

In this section we analyze the multipole structure of the velocity field.
Nevertheless, we start from comparing the coefficients $C_i$ of the generalized
Tully-Fisher relation to that obtained earlier by \citet{ref:ParGayd05}. They
changed not very significantly; one should note the decrease of the coefficient
corresponding to the morphological type of the galaxy and a slight increase of
the coefficient corresponding to the blue diameter. However, these trends are
also present in the non-relativistic model and are caused by slightly different
statistical properties of the updated sample. It is also interesting that the
difference of the main coefficients $C_1$ for the semirelativistic models with
fixed $\gamma$ and with free $\gamma$ appeared to be 22 per cent, which is
consistent with the 18 per cent difference in distances for these two models
and is naturally caused by the same reasons.

For each regressor we calculated not only the coefficient and its error but also
its statistical significance according to Fisher test. For the semirelativistic
model with fixed $\gamma$ for the subsample with $R_{max}=10000\,\rmn{km\,s}^{-1}$
the minimum value $F=11.5$ corresponds to $C_1$, the maximum value $F=109.7$ --
to $C_2$. These values should be compared to the values $3.8$, $6.6$, $7.9$,
$10.8$ and $12.1$, which correspond to $95$, $99$, $99.5$, $99.9$ and $99.95$
per cent confidence levels respectively. Thus, all the coefficients of the
generalised Tully-Fisher relation (\ref{eqn:TFR}) are statistically significant
at the $99.9$ per cent confidence level.

\begin{figure}
\begin{minipage}{84mm}
\includegraphics[width=84mm]{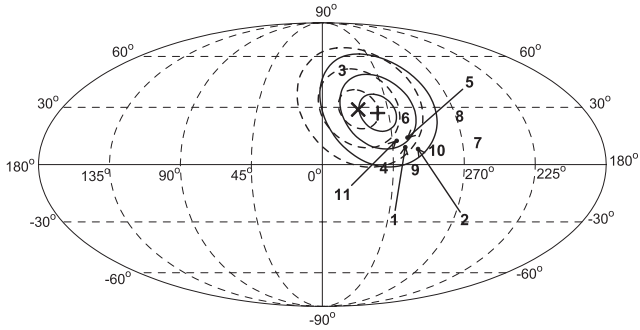}
\caption{Bulk motion apices in galactic coordinates (Mollweide projection) for
$R_{max}=10000\,\rmn{km\,s}^{-1}$.
Crosses mark the apices of the bulk motion in the D-model surrounded by
$1\sigma$, $2\sigma$ and $3\sigma$ confidence areas. Solid boundaries correspond
to results in the relativistic D-model with fixed $\gamma$, and dashed ones -- to
the results of the non-relativistic D-model \citep{ref:APSS09}. Numbers denote the
results of other authors: 1 -- \citep{ref:LB}, 2 -- \citep{ref:H95},
3 -- \citep{ref:LP}, 4 -- \citep{ref:Par01}, 5 -- \citep{ref:Dekel99},
6 -- \citep{ref:daCosta00}, 7 -- \citep{ref:SMAC04}, 8 -- \citep{ref:Dale99},
9 -- \citep{ref:Kudrya03}, 10 -- \citep{ref:WFH09}, 11 -- \citep{ref:ParTug04}}
\label{fig:3}
\end{minipage}
\end{figure}

Now let us consider the dipole component of the velocity field. Its parameters
including the galactic coordinates $l, b$ of the apex for the DQO-model are
given in Table \ref{tbl:1}. The norms of the dipolar component do not
contradict the $\Lambda$CDM model. For the model with fixed $\gamma$ the module
of the dipolar component drops to $180\,\rmn{km\,s}^{-1}$. However, the bulk
motion is usually considered in the framework of the simplest dipole models
when the only characteristics of the velocity field are the modulus and the apex
of the dipole component. In our case of DQO-models the velocity field is
more complex and we cannot attribute the bulk motion solely to the dipole component.
For this reason, to compare our results to the results of other authors we also
calculated the dipole component in the framework of the relativistic (the same
as semirelativistic) D-model with fixed $\gamma$. It yields the bulk flow
velocity of $314\,\rmn{km\,s}^{-1}$ directed towards $l=322\degr, b=27\degr$ (Centaurus).
On Figure \ref{fig:3} we plotted the boundaries of $1\sigma$, $2\sigma$ and $3\sigma$
confidence areas of this apex for $R_{max}=10000\,\rmn{km\,s}^{-1}$. For this purpose
we projected the 8-dimensional ellipsoid of errors into the 3-dimensional space and
then projected it on the celestial sphere. On the same figure we also plotted
the boundaries of the confidence areas of the apex in non-relativistic D-model
\citep{ref:APSS09} as well as positions of apices obtained by different authors.
The value of the bulk motion appears to be larger than for DQO-models. For the
subsample with $R_{max}=8000\,\rmn{km\,s}^{-1}$ it is equal to $285\,\rmn{km\,s}^{-1}$.
We see that D-models provide a result, which is closer to that obtained by
\citet{ref:WFH09}, but still consistent with the $\Lambda$CDM model.

Let us now consider the quadrupole component of the velocity field. What is the
physical sense of the quadrupole component? As one can see from the paper
\citep{ref:Par01}, it can be naturally combined with the Hubble constant. As a
result, we obtain the effective `Hubble constant' depending on direction
\begin{equation}\label{eqn:HC}
H(l,b)=H(1+Q_{ik}n_i n_k).
\end{equation}
Naturally, this effective `Hubble constant' is caused by the large-scale collective
motion on the sample scale. To estimate the value of its anisotropy we found
the eigenvalues and eigenvectors of tensor $Q$. The three eigenvectors are
orthogonal and the sum of three eigenvalues is equal to zero because $\tens{Q}$
is a traceless tensor.

We found the eigenvalues and the eigenvectors of the tensor $\tens{Q}$ for two
considered subsamples. For the subsample with $R_{max}=10000\,\rmn{km\,s}^{-1}$ the
maximal eigenvalue $7.1 \pm 1.7$ per cent corresponds to an axis directed
towards $l=98\degr, b=79\degr$ (Canes Venatici) and the opposite direction
(Phoenix). The minimal eigenvalue $-4.1 \pm 1.7$ per cent corresponds to an
axis directed towards $l=195\degr, b=2\degr$ (Gemini) and the opposite
direction (Sagittarius). The third eigenvalue $-3$ per cent corresponds to an
axis directed towards $l=286\degr, b=11\degr$ (Centaurus-Vela) and the opposite
direction (Andromeda-Lacerta). Comparing these values to the non-relativistic
model \citep{ref:APSS09} one can see that both the eigenvalues and the
directions of the axes changed insignificantly. Nevertheless, the two negative
eigenvalues, which are close to each other, have the opposite order in these
two models. In this sense, the positive axis notably stands out, for which the
effective `Hubble constant' exceeds the mean value by 7 per cent. For the
subsample with $R_{max}=8000\,\rmn{km\,s}^{-1}$ the ellipsoid is three-axial and
essentially differs from the oblate spheroid. The maximal eigenvalue $7.1 \pm
1.7$ per cent corresponds to an axis directed towards $l=102\degr, b=85\degr$
(Canes Venatici) and the opposite direction (Sculptor). The minimal eigenvalue
$-5.3 \pm 1.6$ per cent corresponds to an axis directed towards $l=266\degr,
b=5\degr$ (Vela) and the opposite direction (Cygnus). The third eigenvalue $-2$
per cent corresponds to an axis directed towards $l=356\degr, b=1\degr$
(Sagittarius-Scorpio) and the opposite direction (Auriga-Taurus). These values
are very close to those given by the non-relativistic model. Note that the axes
for both subsamples nearly coincide with the exception of reverse order of
negative eigenvalues for $R_{max}=10000\,\rmn{km\,s}^{-1}$.

We also calculated the statistical significance of these eigenvalues. For
$R_{max}=10000\,\rmn{km\,s}^{-1}$ the maximal eigenvalue has $F=18.2$, which means
that it is non-zero at $99.95$ per cent confidence level, and the minimal
eigenvalue has $F=5.8$, which means that it is non-zero at $97.5$ per cent
confidence level. The similar situation holds for the subsample with
$R_{max}=8000\,\rmn{km\,s}^{-1}$ with Fisher values being $18.3$ and $10.7$
respectively. Additionally, we calculated the total statistical significance of
the quadrupole component. The value $V^{qua}$ with its 5 degrees of freedom
appears to be non-zero at over $99.5$ per cent confidence level according to
F-test.

In the same way we calculated the total statistical significance of the
octopole component. The value $V^{oct}$ with its 10 degrees of freedom appears
to be non-zero at over $99.5$ per cent confidence level according to F-test.
The value $\vec{P}$ with its 3 degrees of freedom appears to be non-zero at
slightly less than $90$ per cent confidence level according to F-test. Unlike
the quadrupole component, the octopole one lacks easily interpretable
characteristics like eigenvector apices. The radial velocity field for
$R=8000\,\rmn{km\,s}^{-1}$ and $R_{max}=10000\,\rmn{km\,s}^{-1}$ in the semirelativistic
model, which includes the octopole component, appeared to be very similar to
that in non-relativistic case, depicted on Fig. 6 in the article
\citep{ref:APSS09}. The most prominent feature of both these velocity fields is
a strong inbound flow coming from the direction opposite to the apex of the
bulk flow.

Thus we obtained that the velocity field in the semirelativistic model with
fixed $\gamma$ is very similar to that in non-relativistic case. The difference
between these velocity fields may become significant when more precise and deep
samples will be available. For the existing sample this similarity yields two
conclusions. The first one is that this similarity justifies the form of the
generalised Tully-Fisher relation for the non-relativistic model. In contrast
to the relativistic model, the non-relativistic model (\ref{eqn:TF}) was
introduced empirically. It includes the term, quadratic in distance, which has
no theoretical substantiations. The relativistic model considered here explains
why such a term needs to be included and what its order is. The second one is
that the external information about the cosmological deceleration parameter,
which we introduced into the model by fixing $\gamma$, is consistent with the
observed parameters of RFGC galaxies.

The bulk flow velocity is the most sensitive characteristic to the selection of the model used.
It can change as much as 1.5 times when the same data is processed with
different models of motion ($314\,\rmn{km\,s}^{-1}$ for D-model, $281\,\rmn{km\,s}^{-1}$
for DQ-model, $181\,\rmn{km\,s}^{-1}$ for DQO-model, and $249\,\rmn{km\,s}^{-1}$ for
DQO-model without vector $\vec{P}$ -- all with fixed $\gamma$ for $100
h^{-1}\,\rmn{Mpc}$). This yields two considerations. First, the bulk flow velocity is
a vulnerable characteristic of collective motion for deep samples. Second, some
authors like \citet{ref:brane} believe that the excessively large values of bulk
flow velocities obtained by some authors are a sufficient reason for
abandoning the $\Lambda$CDM cosmology for more exotic theories like brane
cosmologies. We, however, have a different opinion on this matter, and consider
that such results should be double-checked using different models of collective
motion.

\section{Conclusion}

We applied the relativistic model of motion supplied with the generalised
Tully-Fisher relation (\ref{eqn:TFR}) to the sample of 1623 flat edge-on spiral
galaxies from the RFGC catalogue. The analysis of results prompted us to switch
first to the semirelativistic model, and then to the semirelativistic model
with fixed $\gamma$. The parameters of the collective motion obtained in the
framework of this model appeared to be close to that obtained in the
non-relativistic case. We analysed certain reasons behind the decrease of
$\gamma$ in the semirelativistic model. Evolution of galactic diameters,
selection effects, and difference between isophotal and angular diameter
distances appeared to be inadequate to explain this effect. At the same time,
measurement error in \mbox{H\,{\sc i}} line widths and angular diameters can easily provide
such a decrease. This was illustrated in a toy model, which allows analytical
consideration, and then in the full model using Monte Carlo simulations. The
obtained bulk flow velocity is consistent with $\Lambda$CDM cosmology.

\bsp

\label{lastpage}

\end{document}